\documentclass[pt10,journal=jacsat,manuscript=article,layout=twocolumn]{achemso}
\setkeys{acs}{articletitle = true}

\usepackage{units}
\usepackage{graphicx}
\usepackage{amsmath}
\usepackage{esint}
\usepackage{amssymb}  
\usepackage{achemso}
\usepackage[version=3]{mhchem}
\usepackage{titlesec}
\usepackage{footnote}
\usepackage{booktabs}

\makesavenoteenv{tabular}

\titleformat{\section}[display]{\bfseries}{}{0.0ex}{}[] 
\titleformat{\subsection}[runin]{\bfseries}{}{0.0ex}{}[] 
\titleformat{\subsubsection}[runin]{\itshape}{}{0.0ex}{}[] 

\makeatletter

\let\oldmaketitle\maketitle
\let\maketitle\relax

\newcommand{\onlinecite}[1]{\hspace{-1 ex} \nocite{#1}\citenum{#1}}

\title{Materials Screening for the Discovery of New Half-Heuslers: Machine Learning versus Ab Initio Methods}

\author{Fleur Legrain}
\email{fleur.legrain@cea.fr}
\affiliation[CEA, LITEN]
{CEA, LITEN, 17 Rue des Martyrs, 38054 Grenoble, France}
\author{Jes\'{u}s Carrete}
\affiliation[CEA, LITEN]
{CEA, LITEN, 17 Rue des Martyrs, 38054 Grenoble, France}
\altaffiliation{Current address: Institute of Materials Chemistry, TU Wien, A-1060 Vienna, Austria}
\author{Ambroise van Roekeghem}
\affiliation[CEA, LITEN]
{CEA, LITEN, 17 Rue des Martyrs, 38054 Grenoble, France}

\author{Georg K.H. Madsen}
\affiliation[CEA, LITEN]
{Institute of Materials Chemistry, TU Wien, A-1060 Vienna, Austria}
\author{Natalio Mingo}
\email{natalio.mingo@cea.fr}
\affiliation[CEA, LITEN]
{CEA, LITEN, 17 Rue des Martyrs, 38054 Grenoble, France}    

\makeatother

\usepackage[english]{babel}
\begin{document}

\oldmaketitle

\newpage
\begin{abstract}
\noindent
 
Machine learning (ML) is increasingly becoming a helpful tool in the search for novel functional compounds. Here we use classification via random forests to predict the stability of half-Heusler (HH) compounds, using only experimentally reported compounds as a training set. Cross-validation yields an excellent agreement between the fraction of compounds classified as stable and the actual fraction of truly stable compounds in the ICSD. The ML model is then employed to screen 71,178 different 1:1:1 compositions, yielding 481 likely stable candidates. The predicted stability of HH compounds from three previous high throughput ab initio studies is critically analyzed from the perspective of the alternative ML approach. The incomplete consistency among the three separate ab initio studies and between them and the ML predictions suggests that additional factors beyond those considered by ab initio phase stability calculations might be determinant to the stability of the compounds. Such factors can include configurational entropies and quasiharmonic contributions.  
\end{abstract}

\newpage
\section{Introduction}
Due to their flexible composition and resulting tunable functionalities, half-Heusler (HH) compounds are at the focus of considerable attention\cite{HalfHeuslercompoundsnovelmaterials}. Substantial efforts are currently underway to discover new stable HH compounds, with successful examples in a wide range of fields including topological insulators\cite{TheoreticalsearchforhalfHeuslertopologicalinsulators,Observation,TopologicalRPdBi,HalfHeuslerternarycompoundsasnewmultifunctional}, magnets\cite{BasicsandprospectiveofmagneticHeuslercompounds,Accelerateddiscoveryofnewmagnets,Computationalinvestigation}, thermoelectrics\cite{FindingUnprecedentedly,EngineeringhalfHeusler,RecentprogressofhalfHeusler,HighEfficiencyHalfHeuslerThermoelectricMaterials} or photovoltaics\cite{HalfHeuslercompoundswitha1eV}. However the large pool of possible stable HHs --- in the order of thousands of compounds --- is expensive to screen, both experimentally and computationally. 

Several computational studies have sought to identify new stable HHs using high-throughput (HT) ab initio calculations\cite{FindingUnprecedentedly,SortingStable,Computationalinvestigation}. 
To determine the stability of a hypothetical compound with first principles one must test the stability of the material against all other possible phases including those that could result from the decomposition of the hypothetical compound. 
In addition, the enthalpies of formation should include the vibrational contributions since in many cases the stable phases at finite temperature do not correspond to the ones at 0 K\cite{High-ThroughputComputationofThermalConductivity}. 
This increases the cost of the already expensive calculations by orders of magnitude. Therefore HT calculations of phase stability often neglect entropy contributions, and/or limit the number of competing phases studied\cite{FindingUnprecedentedly,SortingStable,Computationalinvestigation}. The larger the number of compounds investigated, the more pressing the need for these approximations. In addition, kinetic effects are known to play an important role in the ability to synthesize a compound, and may hinder the appearance of the thermodynamically most favorable phase\cite{Spectraldescriptorsforbulkmetallicglasses,Robusttopologicalsurfacestate}. Finally, semi-local density functional theory (DFT) can have limited precision in predicting the stability of the HH compounds \cite{Anovelp-typehalf-Heusler}.

A very different approach to predicting compound stability has been recently proposed in Ref.~\onlinecite{HighThroughputMachineLearningDriven}. This method uses a ML algorithm to determine the probability that a given phase is stable at a particular chemical composition. The algorithm is trained only on experimentally reported compounds, not relying on any computed formation enthalpies or ab initio data of any sort. Ref.~\onlinecite{HighThroughputMachineLearningDriven} applied this method to the full-Heusler compounds, demonstrating an excellent performance in cross-validation, and succeeding in predicting several stable Heusler compositions that had not been previously reported. This ML approach also has its own drawbacks, notably its reliance on the quality of the training set, and the difficulty to extract a deeper understanding of the physics and chemistry behind its predictions. Nevertheless, it may prove an invaluable complement to the ab initio predictions in the search for novel compounds. 

It is therefore very important to know to which extent the ML and ab initio approaches differ in their predictions, and to try to understand the possible reasons for these differences. We carry out such a comparison here. We have chosen the family of the half-Heusler compounds, because of its large number of potential compounds, and the existence of at least three published ab initio HT studies dealing with them\cite{FindingUnprecedentedly,SortingStable,Computationalinvestigation}. In what follows we first describe the pre-existing ab initio HT results, and assess their mutual consistency. Afterwards we describe the ML method and our modification from its original form in order to deal with half-Heusler compounds. We then use the ML approach to screen and evaluate the stability likelihood of 71,178 half-Heuslers. Finally we compare the ML and ab initio results, and discuss the possible reasons for the differences between their predictions.

\section{Previous ab initio HT studies of half-Heusler compounds.}

HHs crystallize in the space group F$\bar{4}3$M (Pearson symbol cF12) with the 1:1:1 composition (XYZ). The HH structure is shown on Figure 1. X, Y, and Z form three interpenetrating face-centred cubic sublattices, occupying the 4a, 4c, and 4b Wyckoff positions, respectively. The X and Z atoms are tetrahedrally coordinated (to the Y atom) and are geometrically equivalent; interchanging them does not change the HH structure. On the other hand, the Y atom is octahedrally coordinated and it is not interchangeable with the X or Z atom without changing the structure of the material. The HH structure can be seen as XY (or YZ) forming a zincblende structure with the Z (or X) atoms filling half of the tetrahedral sites of the zincblende structure.

\begin{figure}
\includegraphics[width=0.6\linewidth]{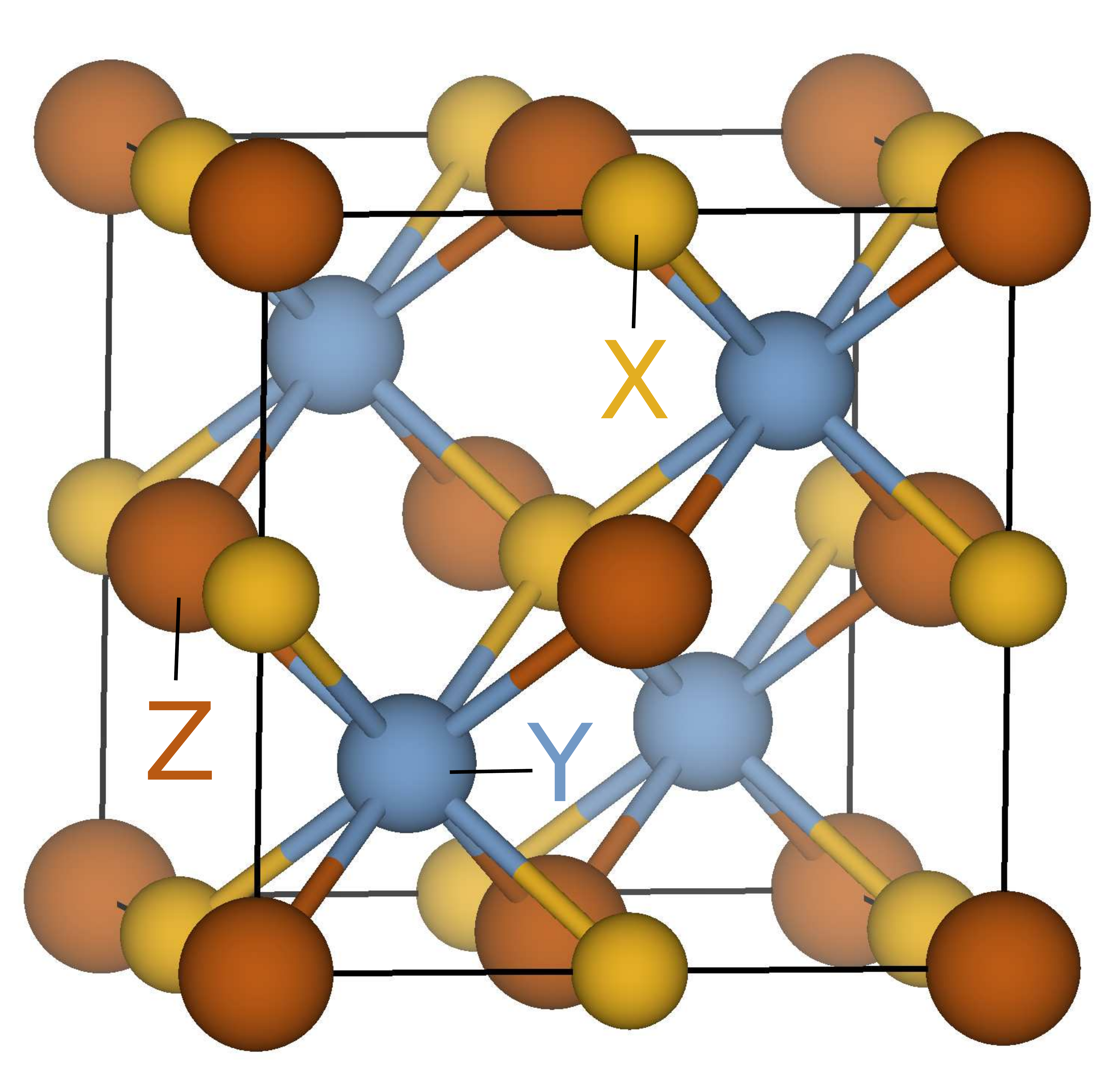}

\caption{HH structure. X and Z sites are geometrically equivalent whereas Y site is different. A chemical composition ABC can form three different HH compounds: A, B, or C occupying the Y site.}

\end{figure}

First let us examine the consistency of the previous HT ab initio studies that explore the stability of hypothetical HH compounds, namely the works of Carrete et al.\cite{FindingUnprecedentedly}, of Zhang et al.\cite{SortingStable}, and of Ma et al.\cite{Computationalinvestigation}. The three HT ab initio studies do not employ the exact same method and do not have the exact same rules to determine the stability of the hypothetical compounds. The results of the ab initio studies are best aligned by setting the same cutoffs of formation enthalpy (0 eV) and of convex hull energy (0 eV). In this way the discrepancies pointed out in the following stem from the different methods and calculations and not from the interpretation of the results.
Because the convex hull energy cutoff is set to 0 eV, only one of the three HH structures that exist per composition (see Figure 1) can be stable. Therefore, for simplicity, the comparison is performed using the composition and not each of the three structures per composition. The composition is said to be \textit{stable} if the most stable HH compound out of the three possible HH structures is stable, and unstable otherwise.
However, for each discrepancy ---i.e. a composition stable by one study but not by the other study--- it is checked that the reported stable HH structure is well explored by the other study (because of resources, the ab initio studies did not systematically consider the three possible structures).

\begin{figure}
\includegraphics[width=0.99\linewidth]{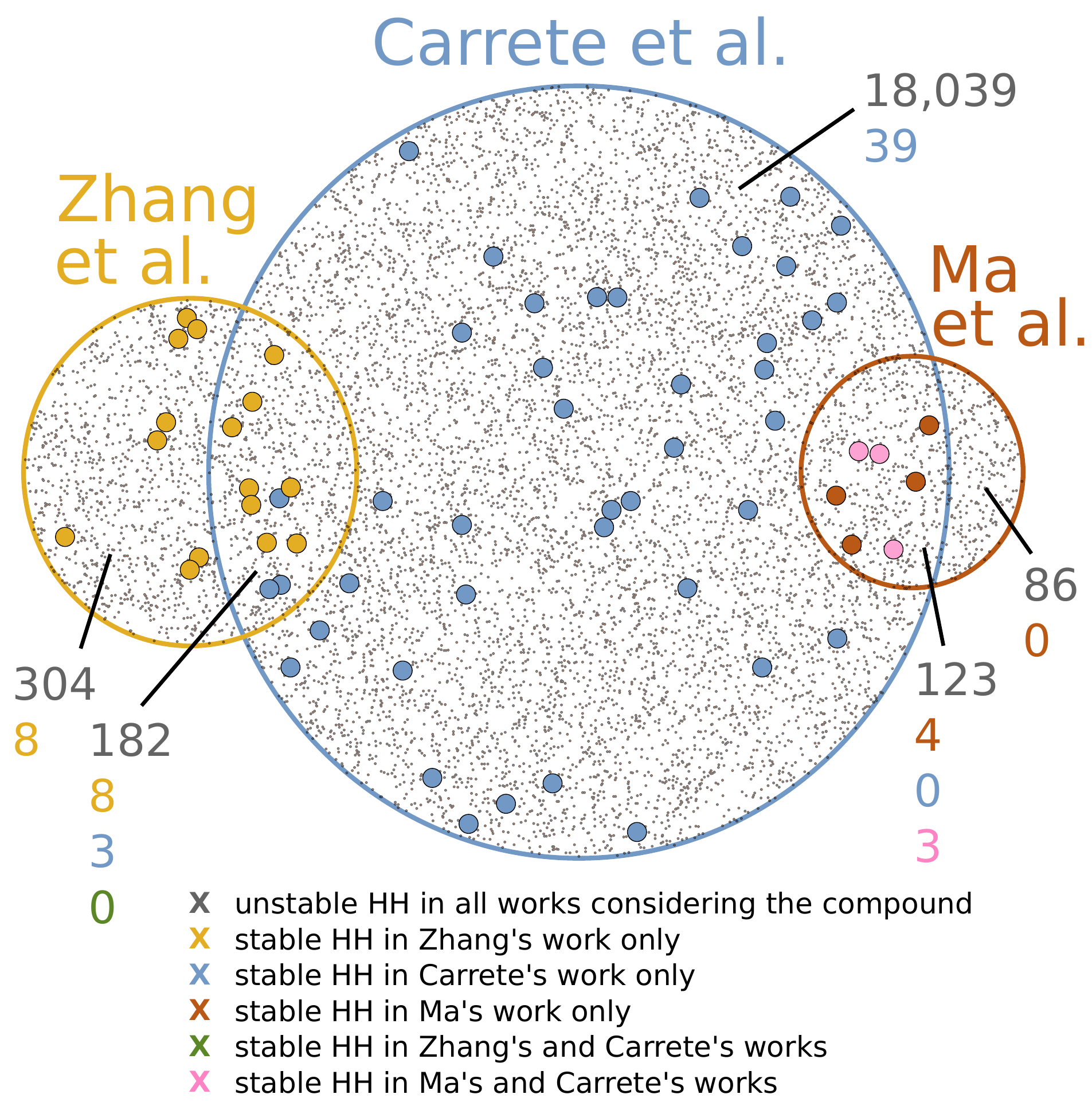}

\caption{The points surrounded by the blue, red, and green circumferences represent the hypothetical HHs considered in Carrete's, Zhang's, and Ma's works, respectively. The intersections between the circles show the overlaps between the sets of compositions explored in the different works. The figures associated with each domain provide the number of compositions of different kinds (unstable / stable, and by which studies). The color scheme is given in the legend.}

\end{figure}

An overview of the interrelation between the three ab initio studies is given in Figure 2. Carrete et al.\cite{FindingUnprecedentedly}, Zhang et al.\cite{SortingStable}, and Ma et al.\cite{Computationalinvestigation} investigated the possible stability in the HH structure of respectively 18501, 505, and 216 different compositions not already reported in the ICSD-aflow.org, either as HH or as some other prototype. As illustrated in Figure 2, 130 compositions are simultaneously contained in the studies
by Carrete and by Ma, and 193 compositions are common to those by Carrete and by Zhang. There is no overlap between the sets considered by Zhang and Ma. Out of the 130 compositions common to Carrete and Ma, 7 HHs (VCoAs, MnCoAs, VRuAs, TiRhSb, TiCoAs, VCoGe, TiNiGe) are found stable by at least one of the two studies, out of which 3 (TiCoAs, VCoGe, TiNiGe) are found stable by the two studies and 4 HHs (VCoAs, MnCoAs, VRuAs, TiRhSb) are found stable in Ma's work but not in Carrete's work. For the 193 compositions common to Carrete and Zhang, 3 compounds (NaSbSr, NaPCd, BaBiK) are found stable by Carrete, and 8 other (CuLiTe, AgLiTe, CuInGe, LiInSi, BeZnSi, BeZnGe, BeZnSn, ZnHgSn) are found stable by Zhang. This means that for the prediction of stable compounds there is no agreement between the two works. To summarize, out of the hypothetical HHs investigated by more than one study (either by Carrete and Zhang or by Carrete and Ma), 18 are found stable by at least one study, out of which only 3 are simultaneously found stable by two studies. The remaining 15 compounds are sources of disagreement between the studies. 

The origin of the discrepancies has several sources. The 3 hypothetical HHs that are found stable in Carrete's work but not in Zhang's work are due to Zhang and coworkers finding another more stable phase: according to Zhang's work BaBiK is more stable in the P63/mmc space group (hP6 Pearson symbol), NaPCd in the Pnma space group (oP12 Pearson symbol), and NaSbSr in the Ima2 space group (oI36 Pearson's symbol). The inconsistencies here can be due to Zhang and coworkers correcting the DFT energies using Ref.~\onlinecite{Correctingdensityfunctionaltheory} and investigating additional competing phases (the phases were not available in the database used by ref.~\onlinecite{FindingUnprecedentedly} at the time, and were not explicitly included because of the computational cost.) The 8 hypothetical HHs that are found stable in Zhang's work but not in Carrete's work are so because of different reasons: Carrete's calculations gave positive formation energies for 5 HHs (CuInGe, BeZnSi, BeZnGe, BeZnSn, ZnHgSn), mechanical unstability for 1 HH (AgLiTe), and thermodynamic instability vs. other competitive phases for the 2 remaining HHs (CuLiTe, LiInSi). For the 4 hypothetical HHs that are found stable in the work of Ma but not in the work of Carrete, the reason is a mechanical unstability for 2 HHs (VCoAs, MnCoAs) and a thermodynamical unstability vs. other competitive phases for the other 2 HHs (VRuAs, TiRhSb). For those last 8+4 HHs, the inconsistencies can be due to uncertainties in calculations (never out of the question in high-throughput studies), to approximations (in particular of not checking the mechanical stability or of not considering some competitive phases for the thermodynamical stability), or to energy discrepancies (which can sometimes be of just a few meV).

\section{Machine Learning approach}

We use a random forests algorithm\cite{RF} that employs a number of trees (set to 1,000 here) to make the stable / unstable classification. 
In particular, every tree classifies each of the inputs.
In what follows, the trees classifying the input as stable are referred as \textit{positive trees} and those classifying the input as unstable as \textit{negative trees}. The fractions of positive and negative trees are computed by the ML model. The classification (stable / unstable) of the input is based on the majority trees. The descriptors employed are based on the elemental properties of the three atoms of the compounds. In order to achieve better performance, the geometrically equivalent X and Z atoms are distinguished. Since the electronegativity of the X and Z atoms play an important role in the stability of HH\cite{EngineeringhalfHeusler}, X and Z are discriminated based on their electronegativities. The full list of descriptors employed is given in Table 1. 

\begin{table*}
\caption{List of the full set of descriptors with their relative importance in the ML model. X (Z) is set as the most (least) electropositive atom. We use elemental properties of the atoms X, Y, Z (atomic number \textit{Z}, column number \textit{col}, row number \textit{row}, radius \textit{r}, mass \textit{m}, electronegativity $\chi$) and some of their combinations (specifically, differences and ratios of the radii and electronegativities of the different atoms). The column number of the lanthanides is set to 3. The descriptors also include the covariances between all elemental properties, as they were shown to significantly enhance the predictions\cite{Representationofcompoundsformachinelearning}. The covariance between two elemental properties $i$ and $j$ is indicated as $cov_{i,j}$. Additional descriptors are the numbers of s, p, d valence electrons ($s$, $p$, $d$). We also consider the sum of the column numbers of the three atoms ($sum_{col}$), in order to account for the sum of valence electrons, as well as the modulo 10 of this sum ($sum_{col}^{10}$).}
\centering
\scriptsize
\begin{minipage}{0.25\textwidth}
\begin{tabular}{ccc}
\toprule
  Ranking & Feature & Importance\\
\midrule
1 & $cov_{col,r}$     &       0.0549\\
2 & $cov_{col,\chi}$     &       0.0346\\
3 & $cov_{r,\chi}$     &       0.0337\\
4 & $\chi_X/\chi_Z$     &       0.0283\\
5 & $cov_{col,m}$     &       0.0273\\
6 & $sum_{col}$     &       0.0272\\
7 & $r_Z/r_Y$     &       0.0265\\
8 & $\chi_X-\chi_Z$     &       0.0262\\
9 & $cov_{Z,col}$     &       0.0259\\
10 & $r_Z-r_Y$     &       0.0247\\
11 & $r_X/r_Z$     &       0.0236\\
12 & $m_Z$     &       0.0235\\
13 & $\chi_X$     &       0.0230\\
14 & $cov_{col,row}$     &       0.0227\\
15 & $Z_Z$     &       0.0224\\
16 & $r_X-r_Z$     &       0.0219\\
17 & $cov_{m,\chi}$     &       0.0217\\
\bottomrule
\end{tabular}
\end{minipage} \hfill
\begin{minipage}{0.28\textwidth}
\begin{tabular}{ccc}
\toprule
  Ranking & Feature & Importance\\
\midrule
18 & $\chi_Z/\chi_Y$     &       0.0214\\
19 & $cov_{Z,m}$     &       0.0206\\
20 & $cov_{Z,\chi}$     &       0.0203\\
21 & $cov_{row,\chi}$     &       0.0197\\
22 & $\chi_X/\chi_Y$     &       0.0197\\
23 & $cov_{r,m}$     &       0.0195\\
24 & $\chi_Z-\chi_Y$     &       0.0193\\
25 & $cov_{Z,r}$     &       0.0188\\
26 & $r_{avg(XZ)-r_Y}$     &       0.0187\\
27 & $\chi_X-\chi_Y$     &       0.0186\\
28 & $col_Y$     &       0.0186\\
29 & $col_Z$     &       0.0180\\
30 & $cov_{row,r}$     &       0.0180\\
31 & $Z_X$     &       0.0177\\
32 & $r_Z$     &       0.0175\\
33 & $sum_{col}^{10}$     &       0.0171\\
34 & $r_X/r_Y$     &       0.0170\\
\bottomrule
\end{tabular}
\end{minipage} \hfill
\begin{minipage}{0.28\textwidth}
\begin{tabular}{ccc}
\toprule
  Ranking & Feature & Importance\\
\midrule
35 & $cov_{row,m}$     &       0.0168\\
36 & $r_X-r_Y$     &       0.0165\\
37 & $d$     &       0.0163\\
38 & $m_X$     &       0.0160\\
39 & $cov_{Z,row}$     &       0.0159\\
40 & $\chi_Z$     &       0.0155\\
41 & $p$     &       0.0152\\
42 & $r_Y$     &       0.0149\\
43 & $\chi_Y$     &       0.0144\\
44 & $r_X$     &       0.0140\\
45 & $Z_Y$     &       0.0133\\
46 & $m_Y$     &       0.0133\\
47 & $col_X$     &       0.0122\\
48 & $row_Z$     &       0.0048\\
49 & $row_Y$     &       0.0043\\
50 & $s$     &       0.0042\\
51 & $row_X$     &       0.0037\\
\bottomrule
\end{tabular}
\end{minipage}
\end{table*}

The data set employed for the training and validation consists of the compounds with the 1:1:1 (ABC) composition that are reported in the Inorganic Crystal Structure Database (ICSD) available in the aflow.org\cite{AFLOWAnautomaticframework,AFLOW104,Aforexchangingmaterialsdata} repositories. For each 1:1:1 composition present in the ICSD-aflow.org, if one of the reported materials is a HH compound, the corresponding stable entry is added to the data set (with the correct correspondence between ABC and XYZ). If not, six unstable entries (for all possible bijections between ABC and XYZ) are added to the data set. In addition, duplicates are removed. This results in a data set of 164 stable entries and of 11,022 unstable entries.

The performance of the ML model is assessed by carrying out and averaging 10 different 10-fold cross-validations. This method evaluates the performance of the ML model on a data set not included in the training set.  For the prediction of stable HHs, we obtain a precision of 0.91 (out of the 92 HHs predicted to be stable, 84 are truly stable) and a recall of 0.51 (out of the 164 truly stable HHs, 84 are predicted to be stable). There is a small proportion of stable HHs as compared to the unstable HHs, but for the prediction of unstable HHs the recall and precision are superior to 0.99. The performance of the ML model is evaluated differently in Figure 3(a) where we plot the fraction of truly stable HHs against the fraction of positive trees. The plot shows that the fraction of positive trees really reflects the actual probability of the HHs to be stable. The ML model is therefore very helpful to guide further detailed work in the search for new stable HH compounds. Figure 3(b) shows the relative densities of the targets of the data set (stable / unstable) as a function of the fraction of positive trees. The plot shows that the overwhelming majority of unstable HHs are well classified. However, it also shows that part of the stable HHs are not captured by the ML model. 

\begin{figure}
\includegraphics[width=0.99\linewidth]{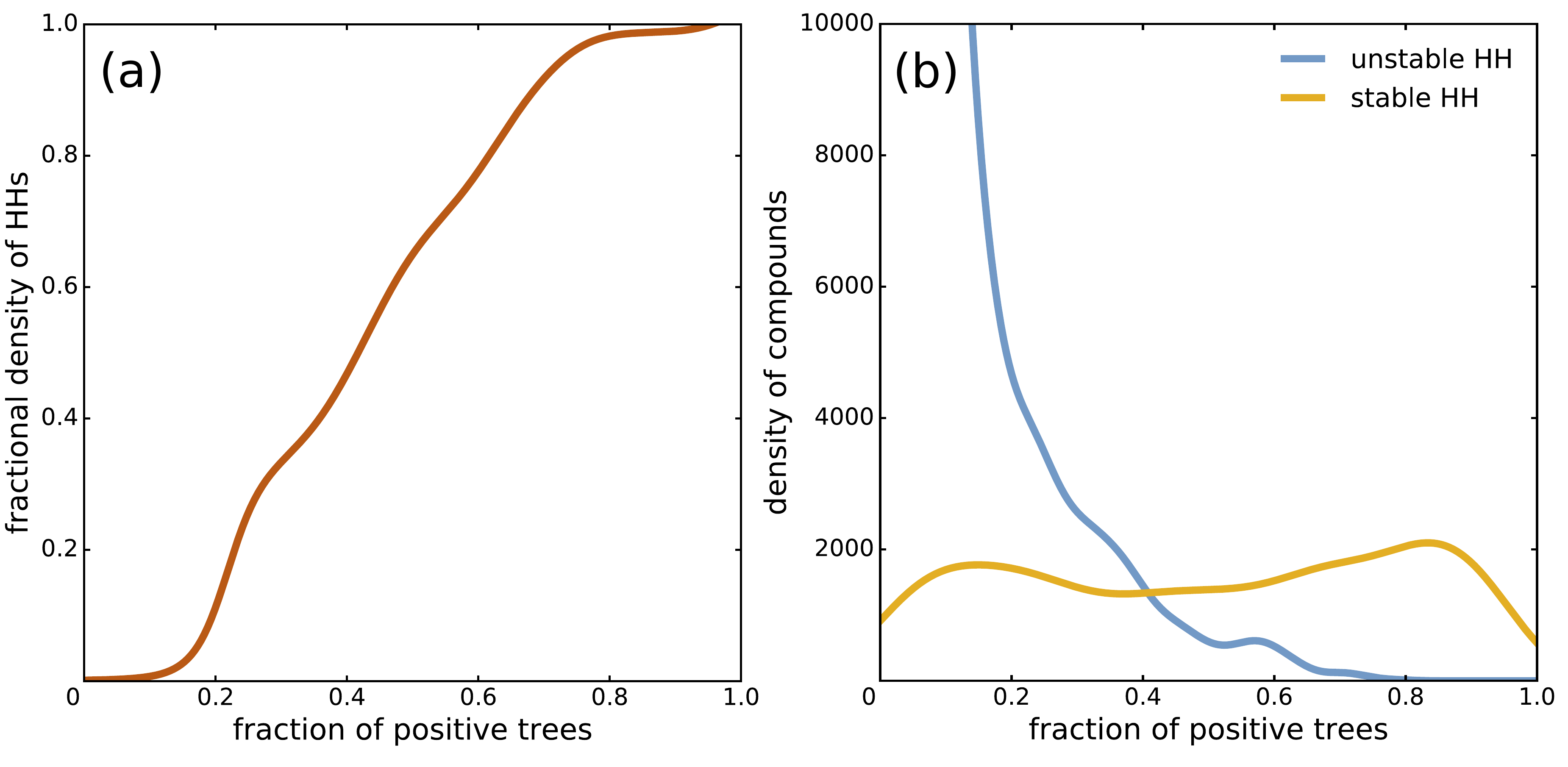}

\caption{Plot of the outcome of the 10-fold cross-validations, i.e. for the predictions of compounds not included in the training set. The y axis shows the targets (the classification as listed in the ICSD-aflow.org) and the x axis shows the predictions (the fraction of positive trees provided by the ML model). (a) the ratio of the density of HHs over the density of all compounds ; (b) the densities of stable and unstable HHs. To plot (a) and (b), the kernel density estimation (KDE) is used.}

\end{figure}

Table 1 gives the relative importance of each feature. It is found that the best ranked features combine elemental properties of the different element types: the most three important features are covariances, of which the covariance between the radii and the column numbers $cov_{r,col}$ comes first. Fig.4 shows the performance of models using smaller sets of features. The features of the subsets are selected by using the recursive feature elimination method with cross-validation. The results show that the subset of 6 features allows for almost the same performance as the full set of descriptors. It contains $r_Z/r_Y$, $\chi_X/\chi_Z$, $cov_{col,r}$, $cov_{col,m}$, $cov_{col,\chi}$, and $cov_{r,\chi}$. The 6 features do not exactly correspond to the best 6 ranked features, due to their intercorrelation. In what follows, the full set of descriptors is used.

\begin{figure}
\includegraphics[width=0.99\linewidth]{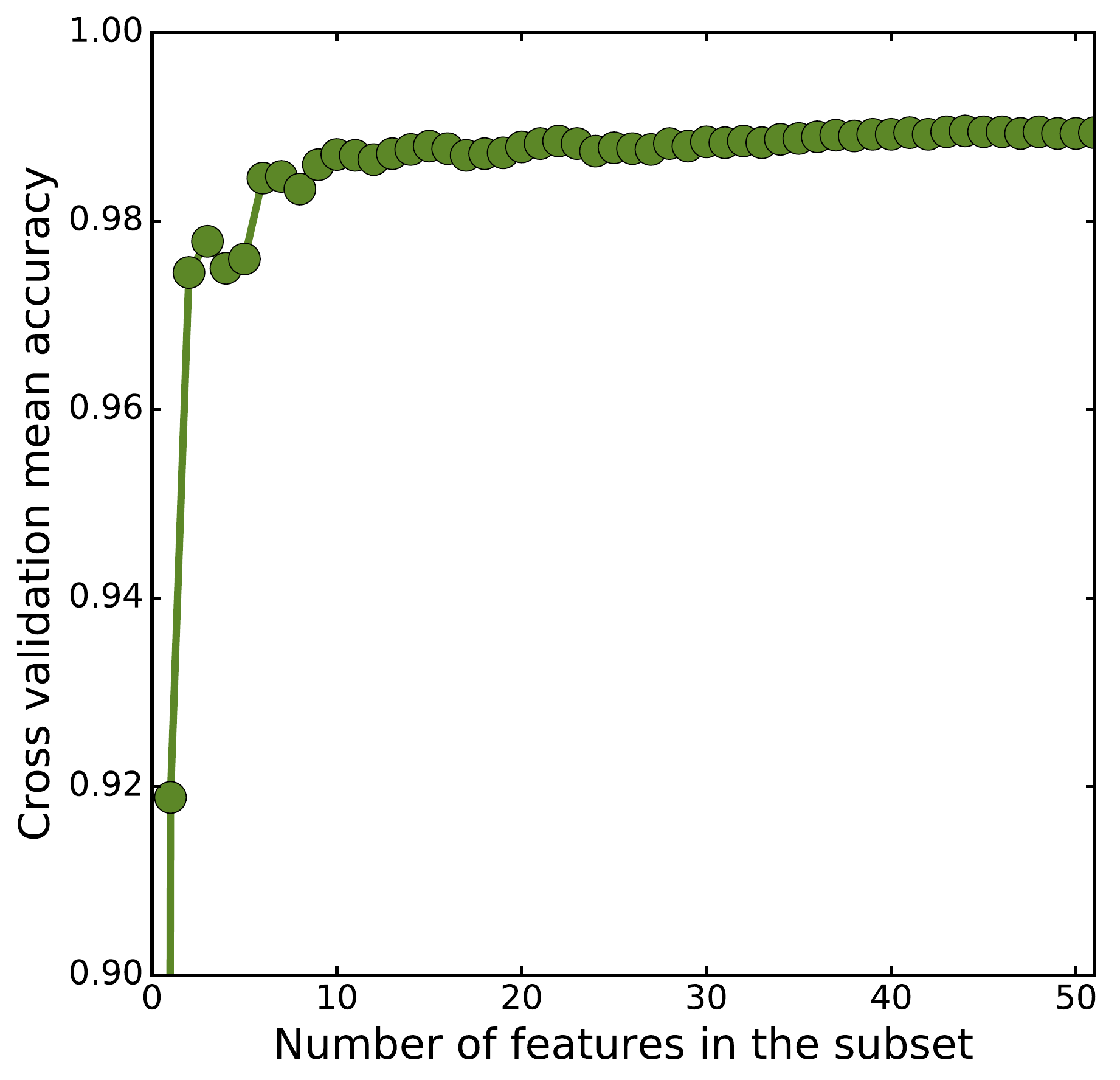}

\caption{The plot shows the score (i.e. the mean accuracy) of subsets that results from 3-fold cross-validation against the number of selected features of the subsets. The features are selected using the recursive feature elimination method.}

\end{figure}

\subsection{Screening for the stability of 71,178 1:1:1 compositions in the HH structure.}

The ML model is used to predict the stability of 1:1:1 (ABC) compositions in the HH structure. The set of study includes all possible 1:1:1 compositions from the elements Li---Bi excluding the noble gases. The compositions already listed in the ICSD (be it in the HH structure or not) are discarded. As a result, 71,178 compositions are explored. For each composition, the 6 different bijections (between A, B, C and X, Y, Z) are considered. For each composition, the results associated with the combination displaying the highest fraction of positive trees are presented. The HHs with a fraction of positive trees superior to 0.8 are given in Table 2, and all compounds with a fraction of positive trees superior to 0.5 are available in the supporting information. In Table 2 and in the supporting information, the stable HHs are given as XYZ, Y being the inequivalent atom. Although the chemical compositions listed in Table 2 and in the supporting information are not present in the ICSD-aflow.org repositories, and thus were absent from the training set, some of them are reported in other databases. In particular a few of them can be found in the Pauling file database\cite{databaseSM}. Many of the hypothetical HHs predicted as most likely stable have actually already been synthesized. Furthermore, two chemical compositions, VCoGe and TiCoAs, that ab initio calculations identify to form in the HH structure but that are experimentally known to exist in another phase\cite{Anovelp-typehalf-Heusler} are correctly classified by the ML model, which provide fractions of positive trees of 0.07 and 0.03 for VCoGe and TiCoAs, respectively. This lends a strong support to the predictive capabilities of the ML algorithm. 

\begin{table}
\centering
\scriptsize
\begin{tabular}{ccc}
\hline
XYZ & Probability & Space group\\
\hline
ErNiBi    &       0.954  & 216$^{a}$\\
TmPtBi    &       0.947  & 216$^{a}$\\
ErPdBi    &       0.931  & 216$^{b}$\\
MnRuSb    &       0.931  & -\\
TbPtBi    &       0.913  & 216$^{a}$\\
TbPdBi    &       0.906  & 216$^{a}$\\
TmPdBi    &       0.899  & 216$^{a}$\\
EuPdBi    &       0.890  & -\\
MnFeSb    &       0.885  & -\\
LuPtBi    &       0.882  & 216$^{a}$\\
YPtBi    &       0.864  & 216$^{a}$\\
EuPtBi    &       0.861  & -\\
TiRhSb    &       0.861  & 216$^{c}$\\
ScPdBi    &       0.854  & 216$^{a}$\\
MnTeRh    &       0.846  & -\\
HfCoBi    &       0.844  & -\\
LuPdBi    &       0.831  & 216$^{a}$\\
MnAgSb    &       0.830  & -\\
TiPtSb    &       0.830  & -\\
TmAuPb    &       0.829  & -\\
VRhSb    &       0.827  & -\\
MnRuTe    &       0.822  & -\\
ZrIrBi    &       0.819  & -\\
MnTePt    &       0.819  & -\\
ScPtBi    &       0.813  & -\\
PmPdBi    &       0.812 & -\\
ZrPdBi    &       0.803  & -\\
MnRhSn    &       0.803  & -\\
\hline
$^{a}$Ref.~\onlinecite{Equiatomic}&        & \\
$^{b}$Ref.~\onlinecite{Thermoelectricandthermophysical}&        & \\
$^{c}$Ref.~\onlinecite{Alloyingbehavior}&        & \\
\end{tabular}
\caption{List of the 28 chemical compositions with a probability of being HH superior to 0.8. The "probability" is expressed as the fraction of positive trees obtained with the ML model. None of these compounds is contained in the ICSD data set we used for the training. However, many of them have been reported in the literature. In those cases the compound's space group number is provided in the third column. Space group 216 is that of the HHs.}
\end{table}

\begin{figure*}
\includegraphics[width=0.99\linewidth]{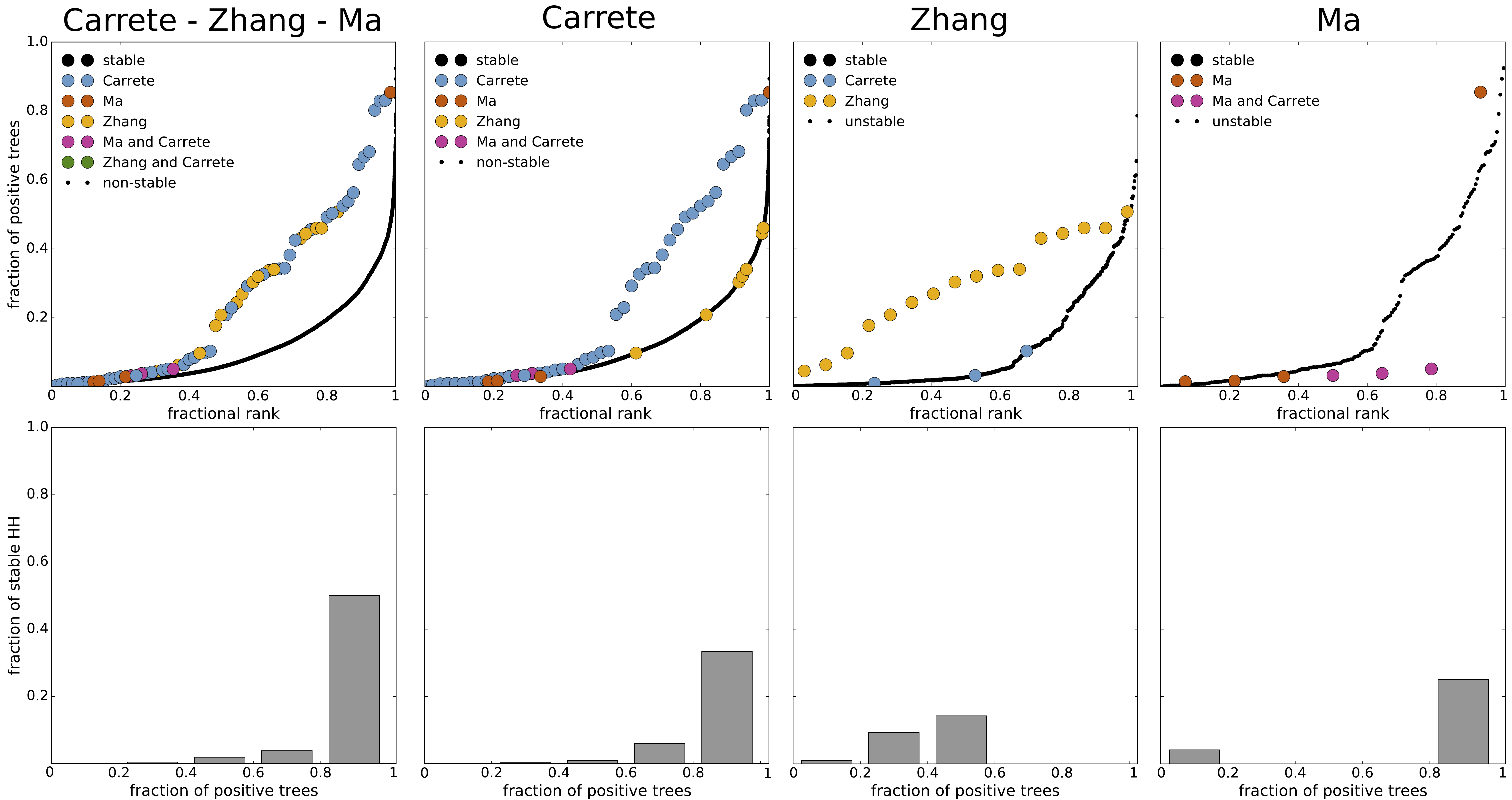}

\caption{Comparison of the predictions from the ML model with those from the ab initio studies. The three ab initio studies (Carrete - Zhang - Ma, on the left) as well as each independent ab initio study (on the middle left, middle right, and right) are used to classify the stable / unstable HHs. For Carrete - Zhang - Ma, a HH is considered as stable if it is found stable by at least one of the three ab initio studies. Top: the y axis is the fraction of positive trees of the hypothetical HHs, the x axis represents the fractional rank of the compound in the stable and unstable lists (when the HHs are ordered in ascending order of their fraction of positive trees). Bottom: the y axis shows the predictions from the ab initio studies (a HH is considered as stable if it is found stable by at least one ab initio study) and the x axis, divided into 5 bins, shows the predictions from the ML model (more specifically the fraction of positive trees).}
\end{figure*}

\subsection{Comparison between ab initio studies and ML model predictions.}
Figure 5 compares the ML predictions with those from the ab initio studies. The histograms on the bottom row of Fig.~5 show the fraction of half-Heusler compounds predicted by the ab initio studies,  as a function of the ML ``probability'', expressed as the fraction of positive trees in the random forests algorithm.
The fraction of stable HHs (according to the ab initio studies) increases with the fraction of positive trees, showing some agreement between the two methods, but the agreement is not as clear as the one observed between ML and experiments (see Figure 2). 

Another way of comparing the two approaches is also provided in Figure 5 (graphs at the top row). The plots show two separate curves, corresponding to the HHs and non-HHs ab initio results. The compounds within each of these two sets are ranked in order of increasing probability, and their probability (fraction of positive trees) is plotted versus this rank number. This yields two monotonically increasing curves. In the ideal case the line of the HH compounds (thick dots) should be concave, meaning that many more compounds would correspond to high ML probabilities. Similarly, the non-HH compound curve should be convex, meaning that the majority of non-HH compounds correspond to low ML probabilities. This depiction then allows for an easy visual evaluation of the agreement between ab initio and ML results.

The actual curve for the ensemble of the three studies indeed shows concave and convex shapes for the HH and non-HH sets respectively. Looking at the separate results of each of the three ab initio studies unveils pattern differences, however. Results of Zhang et al. follow the expected concave-convex pattern quite well, meaning an agreement between the expectations from the ML classification and the actual ab initio results. Carrete's results display the expected convex pattern for the non-HH compounds, but the curve for the HHs is concave only for the higher ranked part of the data, and it has a tail of low-ranked compounds that had been nonetheless classified as HHs by the ab initio calculations. Three of the compounds in this HH tail were also classified as HHs by Ma, and two of them were given as non-HHs by Zhang. For the set of Ma et al., the HH curve is heavily weighed towards the low-probability ML values, opposite to expectation. This unexpected trend is explained by a known DFT failure. DFT calculations identify the XYAs or XYGe compounds to form in the HH structure while experimentally they exist in another structure\cite{Anovelp-typehalf-Heusler}. The six HH compounds that are reported stable in Ma's work but classified as unstable by the ML model (probability inferior to 0.1) are all XYAs or XYGe compounds: VCoAs, MnCoAs, VRuAs, TiCoAs, VCoGe, and TiNiGe. Unlike DFT, ML predicts correctly the stability of those HH compounds. This reconfirms that ML methods are a great help for predicting the stability of HH compounds. The non-HH curve associated to Ma's work roughly follows the expected pattern, but there is an unexpected concave part in the 0.7-0.8 fractional rank region. These qualitative differences in shape between the three ab initio datasets may be due to the different recipes used to choose the chemical compositions of the set. Zhang's set contained compositions with 8 or 18 electron count, whereas those of Carrete and Ma were more diverse. It has also been shown that the zero-kelvin energies computed using DFT were sometimes not sufficient to correctly describe the most stable phase at standard conditions\cite{Anovelp-typehalf-Heusler}. Factors such as configurational entropy and quasiharmonic contributions may also change the ordering of the free energies of competing phases\cite{benisek_vibrational_2015}. In addition, kinetic effects may prevent formation of the thermodynamically most stable phase, in favor of a less stable one that starts nucleating earlier\cite{Robusttopologicalsurfacestate}. Ultimately, the only way to verify the correctness of the ML and ab initio predictions is the experimental verification of the stability as given by the different methods.

\section{Conclusions}

Three different ab initio studies from the literature have provided predictions of potentially new stable HHs. Our analysis of these studies shows that, out
of the 323 compositions for which the ab initio
data sets overlap, 15 hypothetical HHs are found to be stable by one study but not by the other, and 3 hypothetical HHs are found to be stable by the two studies. This suggests that the methodology used today with HT ab initio methods to predict materials stability is not fully consistent among practitioners. Machine-learning algorithms are a powerful complement to ab initio methods, for they are able to guess the stable phase corresponding to a chemical composition by training the model only with experimentally reported compounds. Our ML classification of ternary compositions into half-Heusler versus non-half-Heusler yields an excellent performance on cross-validation, with 91\% of the compounds in the HH group and over 99\% of those in the non-HH group being correctly classified. It is also found that to predict the stability of a hypothetical HH, it is best to use descriptors that combine elemental properties of the different atoms. In particular the covariances between elemental properties are found to be important. We further use the algorithm to sort 71,178 previously unreported ternary compositions in order of increasing likelihood of being a stable half-Heusler, yielding a list of about 30 most likely ones worthy of further study. Some of these compositions were not listed in the ICSD used for the training, but appeared to be stable half-Heuslers in another database, confirming the reliability of the method. We have also shown that there is a certain degree of correlation between the predictions of ML and the ab initio results, with some variability amongst the studies. The ab initio study containing 8 and 18 electron compounds, and the lesser number of compounds, seems to be the one whose results correlate best with the ML prediction. Our results suggest that ML prediction of stable phases can represent a powerful ally of ab initio approaches for the discovery of new materials. Further work, especially in combination with experimental validation, is expected to clarify the extent and limitations of ML in this area.

\section{Supporting Information}
List of the HH compounds for which the ML model gives a probability of being stable that is superior to 0.5.

\section{Acknowledgments}
The work is supported by M-era.net through the ICETS project (DFG: MA 5487/4-1) and
ANR through the Carnot MAPPE project.

\bibliography{bib}

\newpage

\end{document}